\documentclass[a4paper,fleqn]{article}
\usepackage{amsmath}
\begin{document}

\title{\textbf{True and fake Lax pairs: how to distinguish them}}

\author{\textsc{Sergei Sakovich}\bigskip \\
\small Institute of Physics, National Academy of Sciences of Belarus \\
\small sergsako@gmail.com}

\date{}

\maketitle

\begin{abstract}
The gauge-invariant description of zero-curvature representations of evolution equations is applied to the problem of how to distinguish the fake Lax pairs from the true Lax pairs. The main difference between the true Lax pairs and the fake ones is found in the structure of their cyclic bases.
\end{abstract}

\section{Introduction}

In this paper, we use the gauge-invariant description of zero-curvature representations of evolution equations, which was developed in \cite{Sak}, to solve the problem of how to distinguish the fake Lax pairs, introduced in
\cite{CN}, from the true Lax pairs. This approach is algorithmic and exploits only the $x$-part of a studied Lax pair. The main difference between the fake Lax pairs and the true ones turns out to be clearly seen from the structure of their \emph{cyclic bases} (we use the terminology of \cite{Sak} and explain it step by step in what follows).

Note that a more general gauge-invariant description, applicable to zero-curvature representations of any non-overdetermined systems of partial differential equations, was introduced in \cite{M1}, in a very abstract form, and later the explicit expressions for the objects involved were provided in \cite{M2}. The concept of cyclic bases of zero-curvature representations, however, did not appear in \cite{M1,M2}.

In Sections \ref{s2} and \ref{s3}, respectively, we study the true and fake Lax pairs considered in \cite{CN}. In Section \ref{s4}, we discuss some further applications of the cyclic basis technique.

\section{True Lax pairs} \label{s2}

Let us study a true Lax pair first. We take the true Lax pair of the modified KdV equation $u_{t}+u_{xxx}-6u^{2}u_{x}=0$ from \cite{CN}:
\begin{gather}
\psi_{xx} = \left(  u^{2} + \mathrm{i} k \frac{u_{x}}{u} - k^{2} \right) \psi + \frac{u_{x}}{u} \psi_{x} , \label{xlax} \\
\psi_{t} = \left(  -\mathrm{i} k \frac{u_{xx}}{u} - 2 k^{2} \frac{u_{x}}{u} \right) \psi + \left( - \frac{u_{xx}}{u} + 2 u^{2} + 2 \mathrm{i} k \frac{u_{x}}{u} + 4 k^{2} \right) \psi_{x} , \label{tlax}
\end{gather}
where $k$ is an essential (spectral) parameter. Introducing the vector $\Phi$,
\begin{equation}
\Phi =
\begin{pmatrix}
\phi_{1}\\
\phi_{2}
\end{pmatrix}
, \qquad \phi_{1} = \psi , \qquad \phi_{2} = \psi_{x} , \label{phi}
\end{equation}
we rewrite the Lax pair (\ref{xlax})--(\ref{tlax}) as
\begin{equation}
\Phi_{x} = X \Phi , \qquad \Phi_{t} = T \Phi , \label{over}
\end{equation}
where
\begin{equation}
X =
\begin{pmatrix}
0 & 1 \\
u^{2} + \mathrm{i} k \frac{u_{x}}{u} - k^{2} & \frac{u_{x}}{u}
\end{pmatrix}
\label{xmat}
\end{equation}
(the explicit form of the matrix $T$ will not be used in what follows). The compatibility condition
\begin{equation}
D_{t} X = D_{x} T - \left[ X , T \right] \label{zcr}
\end{equation}
of the over-determined linear system (\ref{over}) is the \emph{zero-curvature representation} of the modified KdV equation ($D_{t}$ and $D_{x}$ stand for the total derivatives, the square brackets denote the commutator).

For a generic matrix $X(x,u,u_{x},\ldots,u_{x\ldots x})$ and a generic evolution equation $u_{t}=h(x,u,u_{x},\ldots,u_{x\ldots x})$, we can rewrite the zero-curvature representation (\ref{zcr}) in its \emph{characteristic
form} \cite{Sak}:
\begin{equation}
h C_{u} = \nabla H , \label{cov}
\end{equation}
where $\nabla$ is defined by $\nabla M = D_{x} M - \left[ X , M \right]$ for any matrix $M$, the matrix $H$ is determined by $X$ and $T$, and the matrix $C_{u}$ is the \emph{characteristic} of $X$ with respect to $u$, defined as
\begin{equation}
C_{u} = \frac{\partial X}{\partial u} - \nabla \left( \frac{\partial X}{\partial u_{x}} \right) + \nabla^{2} \left( \frac{\partial X}{\partial u_{xx}} \right) - \nabla^{3} \left( \frac{\partial X}{\partial u_{xxx}} \right) + \cdots . \label{char}
\end{equation}
Under the \emph{gauge transformation}
\begin{equation}
X^{\prime} = S X S^{-1} + \left( D_{x} S \right) S^{-1} , \qquad T^{\prime} = S T S^{-1} + \left( D_{t} S \right) S^{-1} , \label{tran}
\end{equation}
generated by the transformation $\Phi^{\prime} = S \Phi$ with any matrix $S$\ ($\det S \neq 0$), the corresponding transformation of (\ref{cov}) is tensor: $C_{u}^{\prime} = S C_{u} S^{-1}$, $\nabla^{\prime} = S \nabla S^{-1}$, $H^{\prime} = S H S^{-1}$. The \emph{cyclic basis} is the sequence of linearly independent matrices $C_{u} , \nabla C_{u} , \ldots , \nabla^{n-1} C_{u}$, where $n$ is maximal. In the \emph{closure equation} of the cyclic basis,
\begin{equation}
\nabla^{n} C_{u} = a_{0} C_{u} + \cdots + a_{n-1} \nabla^{n-1} C_{u} , \label{clos}
\end{equation}
the coefficients $a_{0} , \ldots , a_{n-1}$ (and, of course, the order $n$) are invariants with respect to the transformation (\ref{tran}).

Returning to the particular matrix $X$ (\ref{xmat}), we compute $C_{u}$ and $\nabla^{i} C_{u}$, $i=1,2,3$, and find that $C_{u}$, $\nabla C_{u}$ and $\nabla^{2} C_{u}$ are linearly independent, whereas
\begin{equation}
\nabla^{3} C_{u} = 4 k^{2} \frac{u_{x}}{u} C_{u} + 4 \left( u^{2} - k^{2} \right) \nabla C_{u} + \frac{u_{x}}{u} \nabla^{2} C_{u} . \label{ctru}
\end{equation}
We see that, in the case of the true Lax pair of the modified KdV equation, the cyclic basis is three-dimensional, with the closure equation given by (\ref{ctru}). It is very important that some coefficients of the closure equation contain a parameter: from one hand, this means that the parameter $k$ cannot be removed (`gauged out') from $X$ by (\ref{tran}), since the coefficients of (\ref{ctru}) are invariants, and, from the other hand, this has an essential influence on the structure of the class of all evolution equations which admit the linear problem (\ref{over}) with this particular $X$, as we will see now. Let us find all those evolution equations $u_{t} = h(x , u , u_{x} , \ldots , u_{x \ldots x})$ which are represented by (\ref{over}), where $X$ is given by (\ref{xmat}) and $T(x , u , u_{x} , \ldots , u_{x \ldots x})$ is an arbitrary traceless $2 \times 2$ matrix. Decomposing $H$ in (\ref{cov}) over the cyclic basis as $H = p C_{u} + q \nabla C_{u} + r \nabla^{2} C_{u}$ and using (\ref{ctru}) for $\nabla^{3} C_{u}$, we find
\begin{equation}
q = - \left( D_{x} + \frac{u_{x}}{u} \right)  r , \qquad p = \left( D_{x}^{2} + D_{x} \frac{u_{x}}{u} + 4 \left( k^{2} - u^{2} \right) \right) r \label{coef}
\end{equation}
and
\begin{equation}
h = \left( A + 4 k^{2} B \right) r , \label{exph}
\end{equation}
where
\begin{equation}
A = D_{x}^{2} u^{-1} D_{x} u - 4 D_{x} u^{2} , \qquad B = u^{-1} D_{x} u . \label{ops}
\end{equation}
Taking into account that $r$ can contain only finite-order derivatives of $u$ and that $h$ must be independent of $k$, we find from (\ref{exph}) that $r$ is a polynomial in $k$ and that the general expression for $h$ is $h = R^{n} 0$, $n=1,2,3,\ldots$, where
\begin{equation}
R = A B^{-1} = D_{x}^{2} - 4 u^{2} - 4 u_{x} D_{x}^{-1} u \label{rec}
\end{equation}
is the recursion operator and $D_{x}^{-1} 0$ must be interpreted as any constant. We have obtained a \emph{discrete class} of evolution equations, namely, the integrable hierarchy of the modified KdV equation, whereas a zero-curvature representation without any parameter in the closure equation of its cyclic basis always leads to a \emph{continual class} of evolution equations \cite{Sak}.

\section{Fake Lax pairs} \label{s3}

Now, let us study the fake Lax pair of the modified KdV equation, found in \cite{CN}. Again, we need only the $x$-part of this Lax pair,
\begin{equation}
\psi_{xx} = \left( u^{2} + \mathrm{i} k \frac{u_{x}}{u} + k^{2} \right) \psi + \left( \frac{u_{x}}{u} - 2 \mathrm{i} k \right) \psi_{x} , \label{xfak}
\end{equation}
which is remarkably similar to (\ref{xlax}). Using (\ref{phi}), we rewrite (\ref{xfak}) as $\Phi_{x} = X \Phi$, where
\begin{equation}
X =
\begin{pmatrix}
0 & 1 \\
u^{2} + \mathrm{i} k \frac{u_{x}}{u} + k^{2} & \frac{u_{x}}{u} - 2 \mathrm{i} k
\end{pmatrix}
. \label{mfak}
\end{equation}
For this $X$, we compute the characteristic $C_{u}$ (\ref{char}), and then find that
\begin{equation}
\nabla C_{u} = 0 . \label{fclo}
\end{equation}
Thus, in this `fake' case, the cyclic basis is one-dimensional, and the closure equation (\ref{fclo}) contains no parameters. As a result of this, any evolution equation of the form $u_{t} = D_{x} p(x , u , u_{x} , \ldots , u_{x\ldots x})$, where $p$ is arbitrary, admits a zero-curvature representation (\ref{zcr}) with $X$ (\ref{mfak}), as we can show by decomposing $H$ in (\ref{cov}) over the cyclic basis as $H = p C_{u}$. (Note that the class of represented equations may be even wider if we take into account the \emph{singular basis} \cite{Sak}, but we will not consider this possibility here.) No discrete hierarchy and no recursion operator appeared in this case: we obtained a continual class of evolution equations, as should be expected from the structure of the cyclic basis.

In conclusion, let us study the most general fake Lax pair from \cite{CN}. We rewrite its $x$-part (see (31a) in \cite{CN}) in the matrix form $\Phi_{x} = X \Phi$ with
\begin{equation}
X =
\begin{pmatrix}
0 & 1 \\
\lambda f^{2} + \eta \mu f - \eta^{2} + \eta \frac{D_{x} f}{f} & \mu f - 2 \eta + \frac{D_{x} f}{f}
\end{pmatrix}
, \label{genf}
\end{equation}
where $\lambda$, $\eta$ and $\mu$ are arbitrary parameters and $f(t , x , u , u_{x} , \ldots , u_{x \ldots x})$ is any function. Then we compute $C_{u}$ and $\nabla C_{u}$ for this $X$ (\ref{genf}), and find that
\begin{equation}
\nabla C_{u} = \frac{D_{x} \left( E f \right)}{E f} C_{u} , \label{genc}
\end{equation}
where $E$ is the Euler operator, $E f = \partial f / \partial u - D_{x} \left( \partial f / \partial u_{x} \right) + D_{x}^{2} \left( \partial f / \partial u_{xx} \right) - \cdots$. We see that the cyclic basis for this fake Lax pair is one-dimensional again. The structure of the closure equation (\ref{genc}) suggests that the class of evolution equations, represented by (\ref{zcr}) with $X$ (\ref{genf}), is continual: it contains at least all those evolution equations, for which $f$ is a conserved density. Indeed, considering $\left( 1 \times 1 \right) $-dimensional matrices $X$ and $T$, for which $\left[ X , T \right] = 0$, and choosing $X = f$, we see that (\ref{zcr}) is nothing but a conservation law with a conserved density $f$, whereas the closure equation for this $X=f$ is exactly the same as for $X$ (\ref{genf}), namely (\ref{genc}). Moreover, since the structure of the closure equation is invariant under the gauge transformations, the transformation (\ref{tran}) of $X$ (\ref{genf}) with
\begin{equation}
S = \mathrm{e}^{\eta x}
\begin{pmatrix}
1 & 0 \\
\frac{\eta}{f} & \frac{1}{f}
\end{pmatrix}
, \qquad X^{\prime} =
\begin{pmatrix}
0 & 1 \\
\lambda & \mu
\end{pmatrix}
f \label{sxtr}
\end{equation}
reveals the origin of (\ref{genc}).

\section{Discussion} \label{s4}

We summarize that the difference between true and fake Lax pairs, considered in \cite{CN}, is clearly seen from the structure of their cyclic bases. Below we give some comments on further applications of the cyclic basis technique.

Since the dimensions and the coefficients of cyclic bases of zero-curvature representations are gauge invariants, it is very convenient to use these invariants to verify whether a parameter in a given Lax pair is an essential (`spectral') parameter or this parameter can be removed (`gauged out') by a gauge transformation, and whether two given Lax pairs can (or cannot) be related to each other by a gauge transformation \cite{Sak,S94,S99,KSY,S03,S04,S03a,S11,S14}.

If a given Lax pair contains no essential parameter, there is always a continual class of evolution equations associated with the $x$-part of this linear problem, and most of those equations are non-integrable, of course \cite{Sak,S04,S05a}. However, continual classes can also appear for linear problems with essential parameters, and this indicates that two dependent variables in such a spectral problem can be merged into one new dependent variable by a gauge transformation \cite{Sak,S05}.

Last but not least, the cyclic basis technique is very useful to obtain hierarchies and recursion operators for given spectral problems \cite{Sak,S03,S04,S03a,KKS,SS}.

\end{document}